\documentclass[jcp,aip,10pt,twocolumn,superscriptaddress,notitlepage,showpacs,preprintnumbers,amsmath,amssymb]{revtex4}
\usepackage{amsmath}
\usepackage{amssymb}
\usepackage{dcolumn}
\usepackage{graphicx}
\usepackage{bm}

\begin{document}
\title{$[N]pT$ Monte Carlo Simulations of the Cluster-Crystal-Forming Penetrable Sphere Model }
\author{Kai Zhang}
\affiliation{Department of Chemistry, Duke University, Durham, North
Carolina, 27708, USA}
\author{Patrick Charbonneau}
\affiliation{Department of Chemistry, Duke University, Durham, North
Carolina, 27708, USA}
\affiliation{Department of Physics, Duke University, Durham, North
Carolina, 27708, USA}
\date{\today}

\begin{abstract}
Certain models with purely repulsive pair interactions can form cluster crystals with multiply-occupied lattice sites. Simulating these models' equilibrium properties is, however, quite challenging. Here, we develop an expanded isothermal-isobaric $[N]pT$ ensemble that surmounts this problem by allowing both particle number and lattice spacing to fluctuate. It is particularly efficient at high $T$, where particle insertion is facile. Using this expanded ensemble and thermodynamic integration, we solve the phase diagram of a prototypical cluster-crystal former, the penetrable sphere model (PSM), and compare the results with earlier theoretical predictions.   At high temperatures and densities, the equilibrium occupancy $n_{\mathrm{c}}^{\mathrm{eq}}$ of face-centered cubic (FCC) crystal increases linearly. At low temperatures, although $n_{\mathrm{c}}^{\mathrm{eq}}$ plateaus at integer values, the crystal behavior changes continuously with density. The previously ambiguous crossover around $T\sim0.1$ is resolved.
\end{abstract}
 \maketitle

\section{Introduction}
Motivated by the ubiquity of repulsive electromagnetic interactions, most models for matter include a core component that diverges at zero separation, which results in a first-order fluid-crystal transition. 
Except for a few exotic examples, such as certain bosonic systems~\cite{cinti:2010,saccani:2011},  simple fluids of atoms and small molecules thus similarly crystallize. Yet when nano- and micro-scale particles are concerned, more complex models that include bounded repulsive interactions are also conceivable. Soft and mostly empty particles, such as dendrimers, for instance, can sit on top of each other with only a finite energy penalty~\cite{likos:2006}. A soft-core interaction captures the effective coarse-grained nature of the potential of mean force between them~\cite{lenz:2011}, and at relatively low densities these  interactions are reasonably pairwise additive~\cite{klymko:2011}.

Liquid state theory suggests that the phase behavior of such soft-core repulsive interactions falls into one of two categories, depending on the presence or not of negative components in the Fourier transform of the pair potential -- or, roughly speaking, on its steepness~\cite{likos:2001,mladek:2008}. On the one hand, flat potentials with purely positive Fourier transforms undergo reentrant melting. Under compression at low temperatures, the system first crystallizes and then remelts. The Gaussian core model (GCM), which was first proposed  as a model for plastic crystals~\cite{stillinger:1976} and has since been thoroughly studied~\cite{lang:2000,prestipino:2005,ikeda:2011,ikeda:2011b}, exhibits such behavior. On the other hand, steep potentials with negative Fourier components cluster up at high densities. Under compression, these systems form crystals in which each lattice site may be occupied by multiple particles. The transition from fluid to cluster crystals is driven not only by gains in free volume, but also by reducing the overlap energy~\cite{mladek:2008}. The penetrable sphere model (PSM), which was initially proposed to describe the interaction between polymeric micelles~\cite{marquest:1989}, belongs to this second category. Its properties have been explored by density functional theory (DFT), cell theory, kinetic theory, and basic simulations~\cite{likos:1998,schmidt:1999,santos:2005,falkinger:2007,suh:2010}. To complement these two systems, Mladek \emph{et al.} have introduced the generalized exponential model of index $n$ (GEM-$n$), with a pair potential that scales with distance $r$ like $\exp{(-r^n)}$, to interpolate between the GCM ($n=2$) and the PSM ($n=\infty$)~\cite{mladek:2006}. Since then, a number of additional dynamic and thermodynamic studies have looked at the exotic properties of this class of models~\cite{fragner:2007,mladek:2007,moreno:2007,mladek:2008,neuhaus:2011,zhang:2010b,nikoubashman:2011}.

Studying the thermodynamics of cluster crystals by simulations is, however, particularly challenging. When initializing the system, there is a trade-off between configurations  with more lattice sites but fewer overlaps, and configurations with fewer lattice sites but more overlaps. The problem is that once initialized, (reasonably) finite systems cannot find the average lattice site occupancy $n^{\mathrm{eq}}_{\mathrm{c}}$ because no simple ensemble allows for the lattice spacing and occupancy to simultaneously relax~\cite{swope:1992,mladek:2007,mladek:2008}. Incommensurability and free energy barriers get in the way. A free energy scheme based on an extended thermodynamic description is necessary to locate the equilibrium phase, but the approaches used thus far have come at a fairly high computational cost~\cite{mladek:2007,mladek:2008,zhang:2010b}. In this paper, we propose a  more direct and efficient extended ensemble that effectively allows both the particle number $N$ and the box volume $V$ to fluctuate. We use it to solve the equilibrium properties of the canonical PSM.

The plan of this paper is to first revisit the PSM and the basic thermodynamics of cluster crystals
(Sect.~\ref{sect:model}) and to introduce the simulation methodology (Sect.~\ref{sect:method}). After discussing the results (Sect.~\ref{sect:results}), a brief conclusion follows.

\section{Model and Thermodynamics}
\label{sect:model}
The penetrable sphere model (PSM) was first introduced by Marquest and Witten as a prototype for soft matter interactions~\cite{marquest:1989,footnote:1}. It is described by a pair potential
\begin{equation}
\phi(r)=
\left\{
\begin{array}{ccl}
\epsilon &\mathrm{for}&0\leq r \leq \sigma\\
0&\mathrm{for}&  r > \sigma,\\
\end{array}
\right.
\end{equation}
where the particle diameter $\sigma$ sets the unit of length and the barrier height $\epsilon > 0$ sets the unit of temperature $T$ ($\beta\equiv1/T$) with Boltzmann's constant $k_B$ set to unity for notational convenience.  The model can also be seen as the $n\rightarrow\infty$ limit of the GEM-$n$, with pair potential
\begin{equation}
\phi(r)=\epsilon e^{-(r/\sigma)^n}.
\end{equation}

The phase behavior of the PSM has not been accurately determined, but some of its general features are reasonably well understood. Based on liquid state theory arguments, one expects that systems with bounded yet harshly repulsive pair interactions, such as the GEM-$n$ with $n>2$~\cite{likos:2001,mladek:2006}, should form face-centered cubic (FCC) cluster crystals at high $T$ and number density $\rho\equiv N/V$. Its $N_\mathrm{c}$ lattice sites then have an average occupancy $n_\mathrm{c} = N / N_{\mathrm{c}}\geq 1$. Density functional theory (DFT) predicts that $n_\mathrm{c}$ increases linearly with $\rho$ at high $T$~\cite{likos:2001,falkinger:2007}, while at low $T$ the fluid-solid transition approaches the hard sphere limit~\cite{schmidt:1999}. A cell theory treatment further suggests that increasing $\rho$ in the low $T$ regime results in a continuous evolution of the lattice occupancy~\cite{likos:1998,schmidt:1999}, in contrast to the series of first-order isostructural transitions and critical points observed in the GEM-4~\cite{zhang:2010b,neuhaus:2011}.

\subsection{Expanded Thermodynamics}
In order to study the behavior of cluster-crystal formers, we first review and expand a thermodynamic framework for treating  multiple-occupancy lattices. The core problem with simulating these crystals is that even for very large  systems the average lattice site occupancy cannot generally relax to its equilibrium value once the crystal forms a lattice commensurate with the simulation box. It is only possible to introduce or withdraw planes of defects without causing elastic stress, which severely restricts the extent to which $n_\mathrm{c}$ can be tuned.  Inspired by the generalized thermodynamics of systems with vacancies and interstitials~\cite{swope:1992}, phase rules that take into account the possibility that crystals have multiple particles occupying each lattice site have been formulated~\cite{mladek:2007,mladek:2008}. In order to specify an equilibrium phase, Mladek \emph{et al.} introduced an additional pair of thermodynamic variables: $N_{\mathrm{c}}$ and an associated chemical potential-like quantity $\mu_\mathrm{c}$, roughly corresponding to the free energy penalty of inserting a lattice site.

In this expanded formulation, the constrained Helmholtz free energy $F_{\mathrm{c}}(N,V,T,N_{\mathrm{c}})$ has an exact differential form
\begin{equation}
dF_{\mathrm{c}} = -SdT-pdV+\mu dN + \mu_{\mathrm{c}}dN_{\mathrm{c}},
\end{equation}
where the chemical potential $\mu$ , the pressure $p$, and the entropy $S$ are all implicit functions of $n_\mathrm{c}$, and at equilibrium $F(N,V,T)=F_{\mathrm{c}}(N,V,T,N_{\mathrm{c}}^{\mathrm{eq}})$. The constrained Gibbs free energy $G_{\mathrm{c}}(N,p,T,N_{\mathrm{c}}) = \mu N + \mu_{\mathrm{c}}N_{\mathrm{c}}$ has a similar differential form
\begin{equation}
dG_{\mathrm{c}} = -SdT+Vdp+\mu dN + \mu_{\mathrm{c}}dN_{\mathrm{c}},
\end{equation}
and $G(N,p,T)=G_{\mathrm{c}}(N,p,T,N_{\mathrm{c}}^{\mathrm{eq}})$. Because $\mu$ and $\mu_{\mathrm{c}}$ are functions of $N_{\mathrm{c}}$, they do not correspond to the system's equilibrium quantities unless $N_{\mathrm{c}}$ minimizes the overall free energy of the system, i.e.,
\begin{equation}
\left(\frac{\partial F_{\mathrm{c}}}{\partial N_{\mathrm{c}}}\right)_{N,V,T;N_{\mathrm{c}}=N_{\mathrm{c}}^{\mathrm{eq}}}=0;~ \mathrm{or}~
\left(\frac{\partial G_{\mathrm{c}}}{\partial N_{\mathrm{c}}}\right)_{N,p,T;N_{\mathrm{c}}=N_{\mathrm{c}}^{\mathrm{eq}}}
=0.
\end{equation}
Equilibrium thus occurs when the driving force for changing the number lattice sites vanishes, i.e.,  $\mu_{\mathrm{c}}=0$.
It is, however, much more computationally convenient to fix $N_{\mathrm{c}}$ and vary $N$. The free energy minimization is then applied on the per-particle quantities,
$f_{\mathrm{c}}(\rho,T,n_{\mathrm{c}})\equiv F_{\mathrm{c}}(N,V,T,N_{\mathrm{c}})/N$ or
$g_{\mathrm{c}}(p,T,n_{\mathrm{c}})\equiv G_{\mathrm{c}}(N,p,T,N_{\mathrm{c}})/N$, such that
\begin{equation}
\left(\frac{\partial f_{\mathrm{c}}}{\partial n_{\mathrm{c}}}\right)_{\rho,T;n_{\mathrm{c}}=n_{\mathrm{c}}^{\mathrm{eq}}}
=0;~\mathrm{or}~
\left(\frac{\partial g_{\mathrm{c}}}{\partial n_{\mathrm{c}}}\right)_{p,T;n_{\mathrm{c}}=n_{\mathrm{c}}^{\mathrm{eq}}}
=0,
\label{eq:minimization}
\end{equation}
as obtained using a generalized Gibbs-Duhem relation.

\subsection{Response Function}
\label{subsect:response}


This thermodynamic construction provides important physical insights into the equilibrium response functions of cluster crystals. The heat capacity $C_{V}$, bulk modulus (or compressibility) $B$($\equiv\kappa^{-1}$) and coefficient of thermal expansion $\alpha$, for instance, can be described as having two different physical contributions. The first comes from ``normal'' crystal relaxation, i.e., an affine transformation under compression, and the second from clustering. For reasons that will become clearer below, the affine component is also dubbed the virial contribution.

The response function is generally defined as the derivative of a thermal quantity $Y$ (energy, enthalpy, volume, etc.) with respect to a changing parameter $x$ (temperature, pressure, chemical potential, etc.). In cluster crystals, $Y$ itself is a function of $N_{\mathrm{c}}^{\mathrm{eq}}$, which in turn depends on $x$, so
\begin{equation*}
\frac{\partial Y(x,N_{\mathrm{c}}^{\mathrm{eq}})}{\partial x}
=\left(\frac{\partial Y}{\partial x}\right)_{N_{\mathrm{c}}=N_{\mathrm{c}}^{\mathrm{eq}}}
+ \left(\frac{\partial Y }{\partial N_{\mathrm{c}}}\right)_{x;N_{\mathrm{c}}=N_{\mathrm{c}}^{\mathrm{eq}}}\left(\frac{\partial N_{\mathrm{c}}^{\mathrm{eq}}}{\partial x}\right),
\end{equation*}
where the two terms correspond to the virial and clustering mechanisms, respectively~\cite{footnote:2}. The bulk modulus of cluster crystals, for instance,
\begin{align}
B & \equiv -V\left(\frac{\partial p(N,V,T,N_{\mathrm{c}}^{\mathrm{eq}})}{\partial V}\right)_{N,T}=\underbrace{-V\left(\frac{\partial p}{\partial V}\right)_{N,T,N_{\mathrm{c}}=N_{\mathrm{c}}^{\mathrm{eq}}}}_{B_\mathrm{vir}}\nonumber\\
&-V\left(\frac{\partial p}{\partial N_{\mathrm{c}}}\right)_{N,V,T;N_{\mathrm{c}}=N_{\mathrm{c}}^{\mathrm{eq}}} \left(\frac{\partial N_{\mathrm{c}}^{\mathrm{eq}} }{\partial V}\right)_{N,T}
\end{align}
has an additional softening contribution due to clustering. Under compression, the lattice relaxes not only by affinely reducing the lattice constant, but also by eliminating lattice sites and clustering~\cite{mladek:2007,mladek:2008,zhang:2010b}. Analogously, the coefficient of thermal expansion
\begin{align}
\alpha & \equiv \frac{1}{V}\left(\frac{\partial V(N,p,T,N_{\mathrm{c}}^{\mathrm{eq}})}{\partial T}\right)_{N,p}\nonumber=\underbrace{\frac{1}{V}\left(\frac{\partial V}{\partial T}\right)_{N,p,N_{\mathrm{c}}=N_{\mathrm{c}}^{\mathrm{eq}}}}_{\alpha_{\mathrm{vir}}}\\
&+\frac{1}{V}\left(\frac{\partial V}{\partial N_{\mathrm{c}}}\right)_{N,p,T;N_{\mathrm{c}}=N_{\mathrm{c}}^{\mathrm{eq}}} \left(\frac{\partial N_{\mathrm{c}}^{\mathrm{eq}} }{\partial T}\right)_{N,p}
\label{eq:alpha}
\end{align}
reveals that under isobaric heating the dilation of the lattice is accompanied by a break up of the clusters themselves, which consumes some of the additional expansion.

Instead of calculating the thermal derivative, it is well known that response functions can also be expressed as fluctuations of thermal quantities, e.g., $C_V = \beta(\langle E^2\rangle-\langle E\rangle^2)/T$~\cite{kubo:1966,allen:1987}. Although formally also true for cluster crystals, special attention needs to be given to $N_{\mathrm{c}}$ being held fixed in simulations. In that case, the calculated fluctuations only capture the affine contribution from the lattice. In other words, the simulation at fixed $N_{\mathrm{c}}$ does not measure the fluctuation of equilibrium lattices, but of lattices constrained to the specific equilibrium occupancy at one phase point.  The clustering contribution must thus be separately calculated, in order to calculate the correct equilibrium behavior.

\section{Methods}
\label{sect:method}
Traditional simulation methods either fix the number of particles, as in constant $NVT$ or $NpT$ simulations, or the lattice spacing, as in constant $\mu VT$ simulations, which in all cases prevent cluster crystals from relaxing to their equilibrium $n_c$. An ensemble that would allow both $N$ and $V$ to fluctuate may na\"ively seem to resolve the problem, but such $\mu pT$ ensemble would have unbounded fluctuations. Earlier simulation approaches thus employed elaborate free energy schemes in order to minimize $f_{\mathrm{c}}$ and $g_{\mathrm{c}}$~\cite{zhang:2010b} or equivalently to locate $\mu_{\mathrm{c}}=0$~\cite{mladek:2007,mladek:2008}. In these schemes, locating the equilibrium structure requires tens of free energy calculations, each necessitating one to two dozens of independently sampled state points. This high computational cost limits the method's applicability to all but the simplest models. In order to reduce the computational burden and broaden the range of accessible models one should limit resorting to free energy integrations. We present below such an approach based on reformulating the expanded isothermal-isobaric $[N]pT$ ensemble~\cite{orkoulas:2009}.

\subsection{$[N]pT$ Ensemble}
To elucidate the $[N]pT$ ensemble, it is useful to first review the generalized (great grand canonical) ensemble~\cite{guggenheim:1939,hill:1987}. Its trivial partition function
\begin{align}
\Upsilon &=\sum_{N}e^{\beta\mu N}\sum_{V}e^{-\beta p V}\sum_{E}\Omega(N,V,E)e^{-\beta E}\nonumber\\
&=\sum_{N}e^{\beta\mu N}\sum_{V}e^{-\beta p V}Q(N,V,T)\\
&=\sum_{N}e^{\beta\mu N}\Delta(N,p,T)=\sum_{V}e^{-\beta p V}\Xi(\mu,V,T)= 1\nonumber,
\end{align}
derived here using a discrete state (histogram) formalism even for continuous variables, implicitly defines the microcanonical $\Omega$, canonical $Q$, isothermal-isobaric  $\Delta$, and grand canonical $\Xi$ partition functions. As mentioned above, in this $\mu p T$ ensemble no extensive quantity is specified, which results in $N$,$ V$ and $E$ having unbounded fluctuations.

An $[N]pT$ ensemble can solve this last problem by constraining $N$ within bounds $[N_{\min},N_{\max}]$ that (if properly chosen)  enclose $n_c^{\mathrm{eq}}$ at the studied $p$, $T$, and $N_c$. The new partition function
\begin{equation}
\Upsilon^*=\sum_{N=N_{\mathrm{min}}}^{N_{\mathrm{max}}}e^{\beta \widetilde{G}_{\mathrm{N}}}\sum_{V}e^{-\beta p V}\sum_{E}\Omega(N,V,E)e^{-\beta E},
\end{equation}
has weights $\widetilde{G}_{\mathrm{N}}\equiv \widetilde{g}_{\mathrm{N}} N$ that are both explicit (extensive) and implicit functions of $N$ to control the probability of visiting states of varying $N$. These weights  are not known \emph{a priori}, but can be self-consistently determined, as is done for the multicanonical ensemble~\cite{berg:1992}, the expanded ensemble parallel tempering~\cite{lyubartsev:1992}, and Wang-Landau sampling~\cite{wang:2001a,wang:2001b}. 
Once the weights are determined, the joint probability of observing a state with a specific $N$, $V$, and $E$ is 
\begin{equation}
\mathcal{P}(N,V,E) =\frac{1}{\Upsilon^*} \Omega(N,V,E)e^{\beta \widetilde{G}_{\mathrm{N}} - \beta p V - \beta E},
\end{equation}
and the marginal probabilities of observing a state with specific $N$ and $V$ or with a specific $N$ only are, respectively,
\begin{align}
&\mathcal{P}_{\mathrm{NV}}(N,V)=\sum_E \mathcal{P}(N,V,E)\sim e^{\beta \widetilde{G}_{\mathrm{N}} -\beta p V }Q(N,V,T)\nonumber\\
&\mathcal{P}_{\mathrm{N}}(N)=\sum_E \sum_V\mathcal{P}(N,V,E) \sim e^{\beta \widetilde{G}_{\mathrm{N}}}\Delta(N,p,T).
\label{eq:marginalP}
\end{align}

\subsubsection{Simulation Method}
During a simulation, in addition to standard particle and logarithmic volume displacements~\cite{frenkel:2002}, particle insertions ($+$) and removals ($-$) are used. In order to preserve detailed balance, the acceptance ratios are
\begin{align}
&\mathrm{acc}(N,V,E\rightarrow E+\Delta E)=\min\left\{1, e^{-\beta\Delta E}\right\}\\
&\mathrm{acc}(N,V\rightarrow V+\Delta  V,E\rightarrow E+\Delta E)=\nonumber\\
& \min\left\{1, e^{-\beta(\Delta E+p\Delta V)+(N+1)\ln\frac{V+\Delta V}{V}}\right\}\\
&\mathrm{acc}(N\rightarrow N\pm1,V,E\rightarrow E+\Delta E^{\pm})=\nonumber\\
& \min\left\{1,\eta^{\pm} e^{\beta\Delta G^{\pm}-\beta\Delta E^{\pm}}\right\},
\end{align}
where $\Delta G^{\pm} =\widetilde{ G}_{\mathrm{N\pm1}}-\widetilde{ G}_{\mathrm{N}}$, $\eta^{+}=V/(N+1)$, $\eta^{-}= N/V$, and $\Delta E^{\pm}$ is the energy cost of inserting/removing a particle. 

In order to self-consistently determine $\tilde{G}_N$, for each iteration $i$ we obtain a histogram $\mathcal{W}^i(N,V,E)$ that keeps track of the frequency at which each $(N,V,E)$ state is observed. After $i$ iterations,  the normalized histogram gives the probability of observing a state
\begin{equation}
\mathcal{P}^i(N,V,E) =\frac{\mathcal{W}^i(N,V,E)}{\sum_{N,V,E}\mathcal{W}^i(N,V,E)}
\end{equation}
that approximates the density of state
\begin{equation}
 \Omega^i(N,V,E)= \mathcal{P}^i(N,V,E)e^{-\beta \widetilde{G}^i_{\mathrm{N}} + \beta p V + \beta E}
\end{equation}
and the other partition functions
\begin{align}
 Q^i(N,V,T)&=\sum_E \Omega^i(N,V,E)e^{-\beta  E}~\mathrm{and}\\
\Delta^i(N,p,T)&=\sum_V Q^i(N,V,T)e^{-\beta  pV}.
\end{align}
Note, however, that even once the scheme has converged $\Omega$, $Q$ and $\Delta$ all lack a multiplicative constant $\Upsilon^*$, so
the conjugate field that is used for the $(i+1)$th iteration
\begin{equation}
\beta\widetilde{G}_{\mathrm{N}}^{i+1}=-\ln \Delta^i(N,p,T)
\end{equation}
also lacks an additive constant $\widetilde{C}^i$. The iterative procedure is repeated until the marginal distribution
$\mathcal{P}_{\mathrm{N}}(N)$ is flat within $[N_{\min},N_{\max}]$, i.e., $e^{\beta\widetilde{G}_{\mathrm{N}}}\Delta(N,p,T)=\mathrm{constant}$ (Eq.~\ref{eq:marginalP}), which coincides with the convergence criterion  $\widetilde{G}_{\mathrm{N}}^{i+1} = \widetilde{G}_{\mathrm{N}}^{i}+\widetilde{C}^i$ for all $N\in[N_{\min},N_{\max}]$. 

A final simulation  using the converged weights constructs the equilibrium $\mathcal{W}(N,V,E)$. This last $[N]pT$ simulation is conceptually equivalent to running a series of $NpT$ simulations with different $N$'s, where the transition probability of hopping from one to the other is governed by $\widetilde{G}_{\mathrm{N}}$. The weights thus capture the constrained Gibbs free energy $G_{\mathrm{c}}$ of the system up to a constant $C$.  Because the equilibrium occupancy is determined by minimizing $g_{\mathrm{c}}$ with respect to $N$ (Eq.~\ref{eq:minimization}), the unknown constant $C$, which leads to a $C/N$ term, must be obtained by  thermodynamic integration at a given $N_0 \in [N_{\min},N_{\max}]$ (Sect.~\ref{app:B}).

\begin{figure}
\includegraphics[width=3.4in]{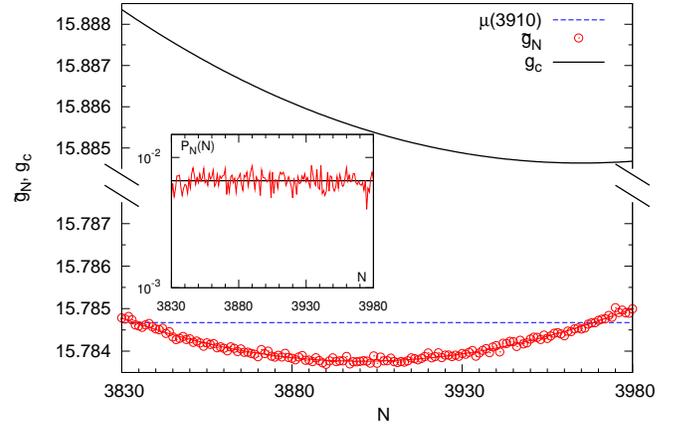}
\caption{$[N]pT$ results for the FCC cluster crystal with $N_{\mathrm{c}}=500$ at $T=1.5$ and $p=33.4$. The weights are initialized with the chemical potential $\mu$ at $N_0 = 3910$ (dashed line) and converge to $\widetilde{g}_{\mathrm{N}}=0.00078065N + 9.6940+11877/N$ ($\odot$). The constrained Gibbs free energy per particle (solid line),  after adding the constant shift $C=397$,  is minimal at $N^{\mathrm{eq}}=3965(10)$. Inset: The converged distribution $\mathcal{P}_{\mathrm{N}}$ is flat within the sampled bounds, fluctuating around 1/151 (black line).} \label{fig:t1.5p33.4}
\end{figure}

\subsubsection{Simulation Details}
For high temperature simulations,  $4\times 10^4$ Monte Carlo (MC) cycles are performed for $N= 1000$--5000 particles at each iteration. Each cycle contains on average $30\% N$ particle number changes, $70\% N$ particle displacements and one volume fluctuation. For the crystal phase, the particles are spread over $N_{\mathrm{c}}=256$--1372 lattice sites depending on the average occupancy. A particle range of $\Delta N \equiv N_\mathrm{max}-N_\mathrm{min}=150$--200 is found to provide a reasonable trade off between the need to correctly estimate the equilibrium occupancy and the computational cost. The histogram of states separates  volume fluctuations in 50 bins and energy fluctuations in 500 bins that span the observed range of fluctuations. For most system sizes studied, the range has $V_{\mathrm{max}}-V_{\mathrm{min}}\sim 50$ and $E_{\mathrm{max}}-E_{\mathrm{min}}\sim 2000$. The summation over these states is done on logarithmic scale to avoid numerical overflow~\cite{newman:1999}.

A good initial guess for the weights is the chemical potential of a system with $N_0 \in [N_{\min},N_{\max}]$, $\widetilde{g}_{\mathrm{N}}=\mu (N_0)$, obtained by Widom insertion in a short $NpT$ simulation~\cite{frenkel:2002}. From this point on, fewer than ten iterations are needed to converge the standard deviation of the equilibrium probability distribution within $\sigma_{\mathcal{P}_N}\simeq 0.001$. Even though the resulting weights are noisy for a given $N$, considering the full $N$ range, by, for instance, fitting them with a parabola smooths out these uncorrelated fluctuations. After including the offset $C$, the constrained Gibbs free energy thus has an approximate form
\begin{equation*}
G_{\mathrm{c}}= c_2 N^2 + c_1N + c_0,
\end{equation*}
where $c_0, c_1$, and $c_2$ are numerically determined. Correspondingly,
\begin{align*}
g_{\mathrm{c}}&=c_2 N + c_1 + c_0/N,\\
\mu &= 2c_2 N +c_1,\;\textrm{and}\\
\mu_{\mathrm{c}}&= -\frac{c_2}{N_{\mathrm{c}}}N^2+\frac{c_0}{N_{\mathrm{c}}}.
\end{align*}
If appropriate bounds are chosen, $g_{\mathrm{c}}$ is then minimized at  $N^{\mathrm{eq}}=\sqrt{c_0/c_2} \in [N_{\min},N_{\max}]$.
Figure~\ref{fig:t1.5p33.4} illustrates this operation. 
Starting with  $\mu(N_0 = 3910)=15.784672$, the simulation converges to $\widetilde{g}_{\mathrm{N}}$. After including the constant offset $C=397$,  $g_{\mathrm{c}}$ is minimal at $N^{\mathrm{eq}}=3965(10)$ ($n_c=7.93(1)$), where the error is obtained from running several optimizations of $\widetilde{g}_{\mathrm{N}}$.  The equilibrium particle number obtained by this method is in good agreement with that determined by pure thermodynamic integration, $N^{\mathrm{eq}}=3960(10)$, following the approach described in Ref.~\cite{zhang:2010b} and Section~\ref{app:B}.
This particular example also illustrates that the minimum of $g_{\mathrm{c}}$ does not need correspond to that of $\widetilde{g}_{\mathrm{N}}$, and can even be out of the chosen
range $[N_{\min},N_{\max}]$, although for the minimization to be reliable within the accuracy reported here the range should be chosen such that the minimum does fall within those bounds.


Determining the equilibrium behavior from the free energy requires a fairly high degree of numerical accuracy. It is thus worth estimating the error made in this process. The typical change in weight values over the full $N$ interval $\Delta g_{\mathrm{c}}$, is smaller than the shift from $\widetilde{g}_{\mathrm{N}}$ to $g_{\mathrm{c}}$. For instance, over an interval of length $\Delta N \sim 10^2$, $\Delta g_{\mathrm{c}}\sim 10^{-3}$. The Gibbs free energy per particle, which is of order $10^2$, is determined by thermodynamic integration with an error $\delta g_0 \sim10^{-2}$. The offset constant $C$ will thus have an error $\delta C=\delta g_0 N_0\sim10^{-2}\times10^3=10^1\sim \delta c_0$. The error $\delta g_{\mathrm{c}}$ in $g_{\mathrm{c}}$ is then of  order $10^{-2}$, which is an order of magnitude larger than the variation $\Delta g_{\mathrm{c}}$. Yet the position of the minimum of the curve, $N^{\mathrm{eq}}=\sqrt{c_0/c_2}$ is only shifted by  $\delta N^{\mathrm{eq}}\sim N^{\mathrm{eq}}(\sqrt{1+\delta c_0/c_0}-1)\sim 10^3\times 10^{-3}=10^0$. Therefore, although the error in the vertical shift of $g_{\mathrm{c}}$ may be an order of magnitude larger than the shallowness of the curve itself, the position of the minimum is preserved with an accuracy comparable to that obtained from running multiple instances of the optimization.

\subsubsection{Low-Temperature Implementation}
Special attention needs to be drawn to use the $[N]pT$ method for the PSM at relatively low temperatures, $T<0.1$. First, the traditional volume fluctuation algorithm, in which the positions of all the particles are affinely rescaled, is inefficient to simulate the PSM. Low temperature compressions of the PSM, like in a system of dense hard spheres, result in additional particle overlaps, whose cost is prohibitively high. To overcome this problem, we use an alternative volume fluctuation algorithm recently developed by Schultz and Kofke~\cite{schultz:2011}, which improves the sampling efficiency by nearly three orders of magnitude at $T=0.05$. In the Schultz-Kofke algorithm, when volume changes from $V$ to $V+\Delta V$ on logarithmic scale, the position of each lattice site $\textbf{R}_i$ is first scaled affinely as $\textbf{R}_{i,\mathrm{new}}=\textbf{R}_{i,\mathrm{old}}\left[(V+\Delta V)/V\right]^{1/3}$, then the distance of each particle from its lattice site $\textbf{r}_j$ is changed as
\begin{equation}
\textbf{r}_{j,\mathrm{new}}=\textbf{r}_{j,\mathrm{old}} \left( \frac{V}{V+\Delta V} e^{\beta (p\Delta V+\Delta E_{\mathrm{lat}})} \right)^{1/(3N-3)}.
\end{equation}
The modified acceptance ratio is then
\begin{equation}
\mathrm{acc}(V\rightarrow V+\Delta V,E\rightarrow E+\Delta E)= \min\left\{1, e^{-\beta(\Delta E-\Delta E_{\mathrm{lat}})}\right\},
\end{equation}
in which the effect of volume change is redistributed into the $\textbf{r}_j$ scaling step~\cite{footnote:3}, and where $\Delta E_{\mathrm{lat}}$ is the change in lattice energy in the $\textbf{R}_i$ scaling step. For step-wise discontinuous potentials like  the PSM, $\Delta E_{\mathrm{lat}}$ is precisely zero if the particles sitting perfectly on lattice sites do not overlap before nor after a volume displacement. Yet if  $\Delta E_{\mathrm{lat}}$ were nonzero, it would scale like $N$, because all the particles on neighboring lattices would go from not overlapping to overlapping. Volume compressions with $\Delta E_{\mathrm{lat}}>0$ are thus rejected with overwhelmingly high probability at low temperatures.

Particle insertion also becomes difficult at low temperatures. For low $T$ crystals with an occupancy intermediate between two integers, new particles can only be successfully inserted at the lower-occupancy positions. Modified particle insertion schemes, such as the staged-insertion algorithm designed for fluid phase~\cite{kaminsky:1994,escobedo:2007}, do not help, because the problem is intrinsic to the system. Inserting particles requires one to randomly identify a vacancy in the crystal, which constrains the $[N]pT$ approach at low $T$ to crystals with non-integer occupancy. Once these regions are identified, PSM crystals with integer occupancy can  then be treated using standard $NpT$ simulations, for instance. For the $[N]pT$ scheme at low $T$, the chemical potential used as initial guess for $\widetilde{G}_{\mathrm{N}}$ is obtained from staged Widom insertion~\cite{mon:1985,kaminsky:1994}, and $10^6$ MC cycles are needed at each iteration to obtain good statistics. To speed up the simulations, more particle insertions and deletions ($50\%N$) and ten to a hundred times fewer volume fluctuations are used in each cycle. The particle number window is also shortened to $\Delta N = 50$. Yet the inherently low efficiency of particle insertion in low $T$ cluster crystals leaves the $[N]pT$ approach less efficient than the free energy integration scheme of Ref.~\onlinecite{zhang:2010b}. Only a few state points were thus equilibrated this way. The rest were obtained with the traditional approach that involves multiple thermodynamic integrations~\cite{zhang:2010b}.

\subsection{Thermodynamic Integration}
\label{app:B}
\begin{figure}
\includegraphics[width=3.5in]{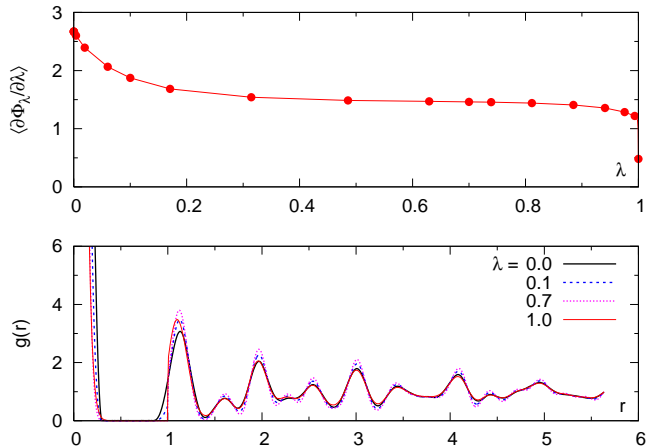}
\caption{Top: Thermodynamic integration of the FCC solid phase with $n_{\mathrm{c}} = 2580/1372 = 1.88$ at $T=0.1$ and $\rho=1.8$. Bottom: The radial distribution function $g(r)=\frac{1}{\rho}\langle\sum_{i=2}^N\delta(\textbf{r}-\textbf{r}_i)\rangle$ at various $\lambda$ shows that the crystal order is preserved along the integration path.} \label{fig:ti0.1}
\end{figure}
Free energy calculation from Kirkwood thermodynamic integration couples the system of interest to a reference system whose free energy can be obtained exactly~\cite{kirkwood:1935,frenkel:2002}. For the cluster crystal, the potential energy of the coupled system with particles at $\textbf{r}^N = \{\textbf{r}_1,\textbf{r}_2,\cdot\cdot\cdot,\textbf{r}_N \}$ is defined as in Ref.~\onlinecite{mladek:2008,mladeck:2007dissertation}
\begin{equation}
\Phi_{\mathrm{\lambda}}=(1-\lambda)\Phi_{0}(\textbf{r}^N)+\lambda\Phi(\textbf{r}^N)
\end{equation}
where
\begin{equation}
\Phi(\textbf{r}^N)=\sum_{i=1}^{N-1}\sum_{j>i}^{N}\phi(r_{\mathrm{ij}}),
\end{equation}
and the coupling parameter $\lambda$ varies from 0 to 1. We choose the reference system to be free particles under spherically attractive potential wells centered at the lattice sites
\begin{equation}
\Phi_0(\textbf{r}^N)=\sum_{i=1}^{N}\phi_0(\textbf{r}_i),
\end{equation}
with
\begin{equation}
\phi_0(\textbf{r})=
\left\{
\begin{array}{ccl}
\epsilon_0 < 0 &,&\textbf{r} \in \bigcup_{i=1}^{N_{\mathrm{c}}} v_0(\textbf{R}_i)\\
0&,& \mathrm{otherwise}\\
\end{array}
\right.,
\end{equation}
where $v_0(\textbf{R}_i)$ is the size of one attractive spherical well at lattice vector $\textbf{R}_i$. The free energy of the reference system at temperature $T$ is therefore~\cite{mladeck:2007dissertation}
\begin{equation}
\beta f_0 = \ln \frac{N}{(V-V_0)+V_0 e^{-\beta \epsilon_0}}-1,
\end{equation}
where $V_0 =N_{\mathrm{c}} v_0$ is the total volume of the potential wells and the thermal wavelength $\Lambda \equiv \sqrt{\beta h^2/2\pi m}$ is a constant at fixed $T$ and is here set to unity. The parameters $v_0$ and $\epsilon_0$ are tuned for each simulation to obtain enough fluctuations and a smooth integration path. The constrained free energy of the solid phase with a certain occupancy is then
\begin{equation}
f_{\mathrm{c}}=f_0+\int_0^1 \frac{\left\langle \Phi-\Phi_0\right\rangle_{\mathrm{\lambda}}}{N} d\lambda,
\end{equation}
as illustrated in Fig.~\ref{fig:ti0.1}~\cite{mladeck:2007dissertation}. It is essential for the integration path to be continuous and tp show no hysteresis, as the preservation of crystal symmetry in Fig.~\ref{fig:ti0.1} also illustrates. For fixed $\rho,T,N_{\mathrm{c}}$, free energy integrations are performed for various $N$'s until the equilibrium $n_{\mathrm{c}}^{\mathrm{eq}}$ is identified from Eq.~\ref{eq:minimization}~\cite{zhang:2010b}.

The free energy of the fluid is obtained from thermodynamic integration from infinite temperature limit where the system is effectively an ideal gas~\cite{frenkel:2002}
\begin{equation}
\beta f = \beta f^{\mathrm{id}} + \int_0^{\beta} \langle \Phi \rangle/N d\beta',
\end{equation}
which is equivalent to setting $\epsilon_0 = 0$ in the crystal integration scheme. Standard integration over density is also used
\begin{equation}
f(\rho_2) = f(\rho_1) + \int_{\rho_1}^{\rho_2} \frac{p}{\rho^2} d\rho.
\end{equation}
to verify the results.

\subsection{Histogram Reweighting}
\label{sec:reweight}
Because  $[N]pT$ ensemble simulations rely on building the full density of state, histogram reweighting provides the thermodynamic properties of the system at neighboring $T$ and $p$~\cite{newman:1999}.

\subsubsection{Reweighting Over Pressure}

Histogram reweighting over pressure, for instance, provides $B_{\mathrm{vir}}$ and $\left(\frac{\partial n_{\mathrm{c}}^{\mathrm{eq}}}{\partial p}\right)_{T}$ from a single $[N]pT$ simulation.
For each $N$ within the bounds, the volume $V$ is sampled with the conditional probability, knowing $N$, at pressure $p$
\begin{equation}
\mathcal{P}_{V|N}(V|N)=\frac{\mathcal{P}_{NV}(N,V)}{\mathcal{P}_{N}(N)}=\frac{e^{-\beta p V }Q(N,V,T)}{\Delta (N,p,T)}.
\end{equation}
The thermal average of $M$ configurations obtained from Monte Carlo sampling at pressure $p$ is thus
\begin{equation}
\langle V \rangle = \frac{\sum_V V\mathcal{P}_{V|N}^{-1} e^{-\beta p V}Q}{\sum_V  \mathcal{P}_{V|N}^{-1} e^{-\beta p V}Q}=\frac{\sum_V V}{M}.
\label{eq:mcvolumep0}
\end{equation}
Now, suppose the simulation is run at a nearby different pressure $p_0$. The probability of observing a state becomes
\begin{equation}
\mathcal{P}_{V|N,0}(V|N)=\frac{e^{-\beta p_0 V }Q(N,V,T)}{\Delta (N,p_0,T)}.
\end{equation}
In order to evaluate the same average quantity $\langle V \rangle$ at $p$, while sampling configurations at $p_0$ subject to the distribution $\mathcal{P}_{V|N,0}$, the histogram is reweighted
 \begin{equation}
\langle V \rangle = \frac{\sum_V V\mathcal{P}_{V|N,0}^{-1} e^{-\beta p V}Q}{\sum_V  \mathcal{P}_{V|N,0}^{-1} e^{-\beta p V}Q}
=\frac{\sum_V V e^{-\beta (p-p_0) V}}{\sum_V  e^{-\beta (p-p_0) V}}.
\end{equation}

\begin{figure}
\includegraphics[width=3.4in]{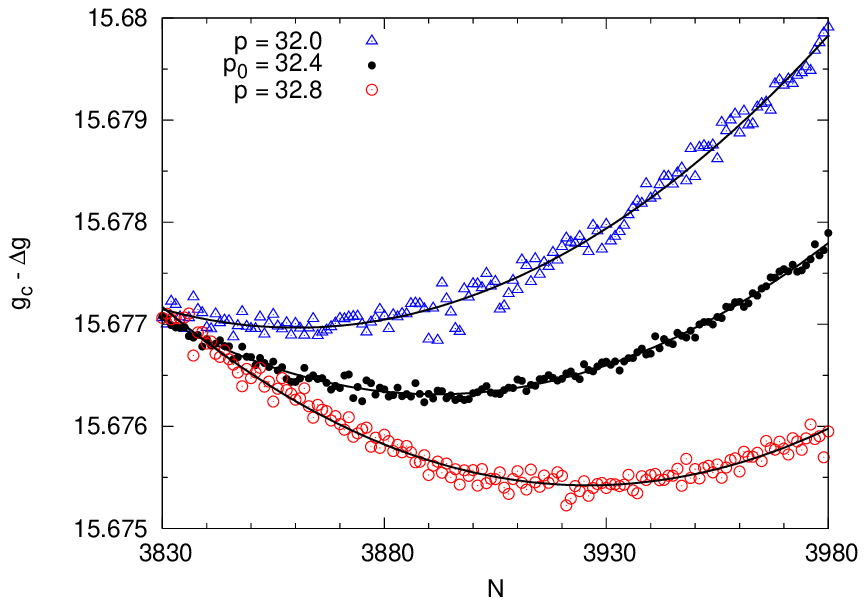}
\includegraphics[width=3.4in]{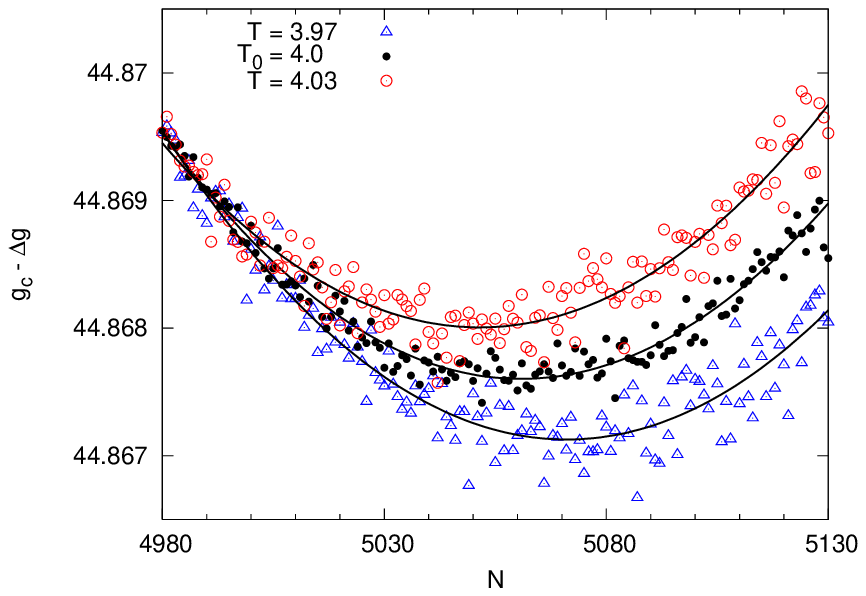}
\caption{Histogram reweighting $g_{\mathrm{c}}$ at (top) $T=1.5$, around $p_0=32.4$, and at (bottom) $p=206$ around $T_0=4$. A constant $\Delta g$ has been subtracted from the free energy in order to superimpose the curves at the beginning of the interval.} \label{fig:reweight}
\end{figure}

For each $n_{\mathrm{c}}$, i.e., for each $N$ at fixed $N_{\mathrm{c}}$, this reweighting scheme provides  $\langle v \rangle = \langle V\rangle/N$ over a short pressure interval enclosing $p_0$.  By integrating  $\langle v \rangle$ for each $n_{\mathrm{c}}$, the constrained Gibbs free energy per particle over a pressure interval follows
\begin{equation}
g_{\mathrm{c}}(p,n_{\mathrm{c}})-g_{\mathrm{c}}(p_0,n_{\mathrm{c}}) = \int_{p_0}^{p}\langle v\rangle dp'.
\label{eq:vintegral}
\end{equation}
For each neighboring pressure $p$, $g_{\mathrm{c}}(p,n_{\mathrm{c}})$ is minimal at  the equilibrium occupancy $n_\mathrm{c}^\mathrm{eq}$ and the equilibrium Gibbs free energy per particle $g(p)$. The corresponding equilibrium Helmholtz free energy per particle $f_\mathrm{c}(\rho) = g_{\mathrm{c}}(p)-p\langle v\rangle$ is similarly available for neighboring densities. Numerically, the volume for each $N$ obtained from $[N]pT$ simulations is reweighted at neighboring pressures and then fitted to a third-order polynomial in pressure, before using Eq.~\ref{eq:vintegral}. Figure~\ref{fig:reweight} shows an example of the resulting  $g_{\mathrm{c}}(p,n_\mathrm{c})$. As a rule of thumb, histogram reweighting is sufficiently accurate as long as $N^{\mathrm{eq}}$ remains within the bounds $[N_{\min},N_{\max}]$.

\subsubsection{Reweighting Over Temperature}

Histogram reweighting over temperature similarly provides  $\alpha_{\mathrm{vir}}$ and $\left(\frac{\partial n_{\mathrm{c}}^{\mathrm{eq}}}{\partial T}\right)_{p}$ from a single $[N]pT$ simulation. One can show that the thermal average of the volume at neighboring temperatures can be calculated from the configurations sampled at $T_0$
 \begin{equation}
\langle V \rangle =\frac{\sum_V  V e^{-(\beta-\beta_0) pV}  \sum_E \Omega(N,V,E) e^{-(\beta-\beta_0) E}}{\sum_V  e^{-(\beta-\beta_0) pV} \sum_E \Omega(N,V,E) e^{-(\beta-\beta_0) E}}.
\label{eq:mcvolumeT}
\end{equation}
To calculate the Gibbs free energy at fixed $p$ and neighboring $T$, one reweights the isobaric-isothermal partition function. Although the real partition function  
differs from the sampling result
by  a multiplicative constant $C^*$, this constant is determined by the free energy at $T_0$ by
\begin{equation}
\beta_0 G_{\mathrm{c}}(N,p,T_0) = - (\ln \Delta(N,p,T_0) +  \ln C^*).
\end{equation}
In a Monte Carlo scheme, the fraction phase space sampled $\Delta(N,p,T_0)$ is proportional to the number of accepted configurations  $M$, because these configurations are already Boltzmann weighted. Note that in this formulation, $C^*$ absorbs the normalization by the total number of the samples considered. For instance, if the algorithm is ran twice as long, $C^*$ is halved. The sampling partition function at neighboring temperatures is thus
\begin{equation}
\Delta(N,p,T) = \sum_{V} e^{-(\beta-\beta_0) pV}  \sum_E \Omega(N,V,E) e^{-(\beta-\beta_0) E},
\end{equation}
and the corresponding constrained Gibbs free energy for each $N$ is
\begin{equation}
\beta G_{\mathrm{c}}(N,p,T) = -  (\ln \Delta(N,p,T) +  \ln C^*).
\end{equation}
Sample reweighting results are shown in Fig.~\ref{fig:reweight}.

\subsection{Phase coexistence}
Identifying the phase coexistence region requires the free energies of the cluster crystal and fluid phases nearby.
For the solid phase, the free energy is obtained at various temperatures using $[N]pT$ simulations, thermodynamic integration,  and histogram reweighting, as described in the
previous sections. The free energy of the fluid phase, which does not suffer from the same occupancy ambiguity, is straightforwardly obtained from standard free energy integration techniques~\cite{frenkel:2002}. For $0.07<T<0.1$ near coexistence, however, standard Monte Carlo sampling of the fluid phase results in long correlations between the configurations. The few particle overlaps present in those conditions do not easily relax because the dense fluid only rarely opens large enough gaps. 
In this regime, longer simulations with $10^7$ MC cycles are necessary to estimate the equilibrium quantities.

The fluid-solid phase coexistence at a specified $T$ is obtained by common tangent construction of the transformed free energy $\widetilde{f}=f\rho-K\rho$. By an appropriate choice of constant $K$, the curvature of the Helmholtz free energy at coexistence can be enhanced~\cite{mladek:2007}. Because
\begin{equation*}
\left(\frac{\partial \widetilde{f}}{\partial \rho}\right)_T =\left(\frac{\partial f}{\partial \rho}\right)_T\rho+f-K=p/\rho+f-K=\mu-K,
\end{equation*}
a common tangent construction also provides the chemical potential and pressure of the two phases concerned. 
%
%
In order to better trace the coexistence line on the phase diagrams, its slope is also calculated at a few points. For the  $p$-$T$ projection, the standard Clausius-Clapeyron relation can be used
\begin{equation}
\left(\frac{\partial p}{\partial T}\right)_{\mathrm{coex}}=\frac{\Delta s}{\Delta v},
\end{equation}
where $\Delta v$ and $\Delta s$ are, respectively, the per particle volume  and entropy difference between the two phases across the transition. The latter quantity is obtained from the free energy results.
For  the $T$-$\rho$ projection, each phase has
\begin{equation}
\left(\frac{\partial\rho}{\partial T}\right)_{\mathrm{coex}} = -\rho\alpha  +\rho\kappa \left(\frac{\partial p}{\partial T}\right)_{\mathrm{coex}}.
\end{equation}
For the fluid branch, the response functions are directly obtained from fluctuation identities~\cite{allen:1987}. The compressibility is  accessible from fluctuations in $N$ for $\mu V T$ simulations, or in $V$ for $NpT$ simulations
\begin{align}
\kappa\equiv-\frac{1}{V}\left(\frac{\partial V}{\partial P}\right)_T&=\beta V\frac{\langle N^2\rangle - \langle N\rangle^2}{\langle N\rangle^2}\nonumber\\
&=\beta\langle V\rangle\frac{\langle V^2 \rangle -\langle V\rangle ^2}{\langle V\rangle ^2},
\label{eq:kappa}
\end{align}
while the coefficient of thermal expansion is estimated from the covariance between $V$ and enthalpy $H=E+pV$ for $NpT$ simulations
\begin{equation}
\alpha\equiv \frac{1}{V}\left(\frac{\partial V}{\partial T}\right)_p=
\frac{\beta}{T}\frac{\langle VH\rangle-\langle V\rangle\langle H\rangle}{\langle V\rangle}.
\label{eq:alphafluc}
\end{equation}
As discussed in Sec.~\ref{subsect:response}, for cluster crystals the fluctuation (virial) contribution is insufficient.
The full derivatives for the solid branch is instead obtained by numerically differentiating the equilibrium results or by reweighting the $[N]pT$ simulation results performed at the equilibrium
pressure and temperature as discussed in Sec.~\ref{sec:reweight}.


\begin{figure*}
\includegraphics[width=2.12in]{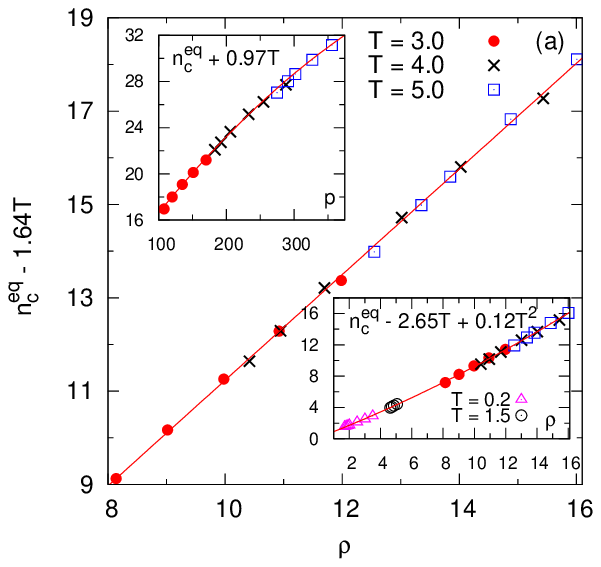} 
\includegraphics[width=2.12in]{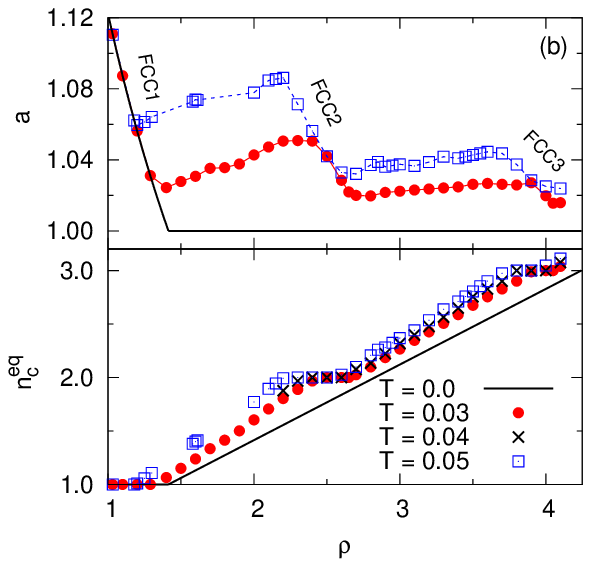}
\includegraphics[width=2.12in]{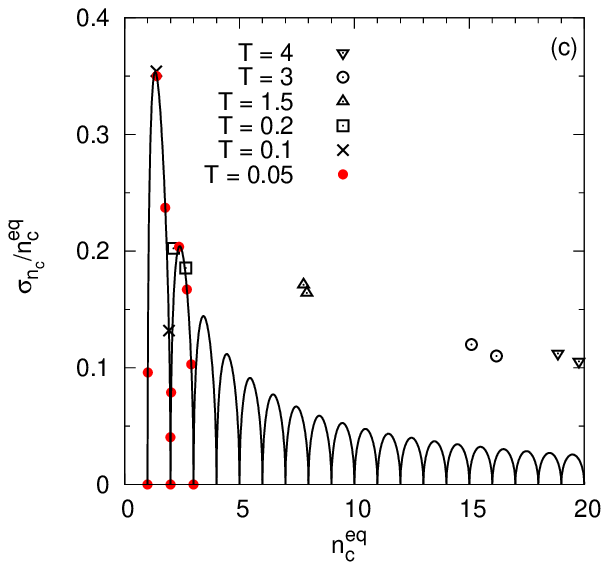}
\caption{(a): Equilibrium occupancy at $T\geq3$ has a bilinear form $n_{\mathrm{c}}^{\mathrm{eq}}(\rho,T)=1.64(3)T+1.132(12)\rho-0.087(108)$. When the lower temperature results are included, higher order terms are necessary to obtain a reasonable fit $n_{\mathrm{c}}^{\mathrm{eq}}(\rho,T)=2.65(11)T-0.12(1)T^2+0.77(5)\rho+0.014(2)\rho^2+0.13(9)$ (right inset) and $n_{\mathrm{c}}^{\mathrm{eq}}(p,T)= -0.000057(7) p^2 + 0.083(4) p + 8.9(4)-0.97(11)T$ (left inset). (b): At low temperatures, the equilibrium occupancy plateaus at integer values (bottom), but the difference in lattice constant  $a=(\sqrt{2} n _{\mathrm{c}}/\rho)^{1/3}$ between neighboring integer occupancy phases is too small to induce demixing (top). At $T=0$, the lattice constant for $\rho>\rho_{\mathrm{cp}}$ is fixed and the occupancy increases linearly.  (c): Relative particle number fluctuation for various temperatures. The theoretical prediction for the fluctuations between $\lfloor n_{\mathrm{c}}^{\mathrm{eq}}\rfloor$ and $\lceil n_\mathrm{c}^{\mathrm{eq}}\rceil$ (solid line --  see text) agree with the low temperature results.} \label{fig:nc}
\end{figure*}

\section{Results and Discussion}
\label{sect:results}
Using the methods described in the previous section the structure and properties of the crystal phase and of the fluid-crystal coexistence of the PSM model are examined.

\subsection{Equilibrium Cluster Crystal Properties}

We first identify the equilibrium cluster crystal occupancy $n_{\mathrm{c}}^{\mathrm{eq}}$ at different phase points. Qualitatively, $n_\mathrm{c}^\mathrm{eq}$ increases with density and pressure, and at fixed pressure the number of particles per cluster decreases with increasing temperature as new lattice sites are formed. 
Quantitatively, at fixed high T, $n_{\mathrm{c}}^{\mathrm{eq}}$  grows roughly linearly with density, while at fixed density, it grows linearly with temperature (Fig.\ref{fig:nc}(a)). If viewed as a function of $\rho$ and $T$,
 $n_{\mathrm{c}}^{\mathrm{eq}}(\rho,T)$ has a bilinear form $n_{\mathrm{c}}^{\mathrm{eq}}=aT+b\rho + c$ with positive $a$ and $b$, and a small $c$. The DFT prediction, $n_{\mathrm{c}}^{\mathrm{eq}}=2T+\rho$~\cite{falkinger:2007}, is qualitatively comparable with the coefficients obtained from fitting the simulation results (Fig~\ref{fig:nc}(a)), but in order to capture the clustering behavior for temperatures as low as $T\sim0.1$, quadratic corrections are necessary.

As $T$ is further lowered, $n_{\mathrm{c}}^{\mathrm{eq}}$ plateaus at integer occupancy values, which we denote FCC($n_c$). At $T=0$, however FCC($n_\mathrm{c}$) with integer $n_\mathrm{c}\geq2$ is
 again only stable at a single density $\rho= n_\mathrm{c} \rho_{\mathrm{cp}}$, where $\rho_{\mathrm{cp}}=\sqrt{2}$ is the crystal close packing density of hard spheres. For $n_\mathrm{c} \rho_{\mathrm{cp}}<\rho<(n_\mathrm{c}+1) \rho_{\mathrm{cp}}$ the
 ground state is a random mixture of sites with occupancy $n_\mathrm{c}$ and $n_\mathrm{c}+1$, because all cluster arrangements are degenerate (Fig.\ref{fig:nc}(b)). 

What about intermediate temperatures? In the GEM-4, transitions between cluster phases with different integer occupancy FCC($n_\mathrm{c}$)--FCC($n_\mathrm{c}+1$) was found to become first order below an isotructural critical point~\cite{zhang:2010b}. Although a simple phonon analysis suggests that a similar type of transition may be possible in any cluster-forming system at low enough temperature~\cite{neuhaus:2011}, no such transition is observed here. This point deserves further consideration. For the small integer occupancy studied, a multiply-occupied lattice site is effectively a large coarse-grained particle whose size is related to the lattice constant $a$.  To obtain a first order isotructural transition, i.e., demixing of FCC($n_\mathrm{c}$) and FCC($n_\mathrm{c}+1$) phases, a large enough size heterogeneity between sites of different occupancy is necessary for the free volume gained to overcome the entropy of mixing, as in binary hard sphere mixtures~\cite{kranendonk:1991a}. Otherwise, the sites with occupancy $n_\mathrm{c}$ and $n_\mathrm{c}+1$ randomly mix. For the PSM, the evolution of the lattice constant in Fig.~\ref{fig:nc}(b) indicates that the largest lattice difference between FCC(1) and FCC(2) is smaller than $3\%$ at $T=0.05$. Because this difference goes to zero at $T=0$, its relatively rapid shrinking with $T$ may be sufficient to prevent isostructural phase separations, although it cannot be excluded altogether. This qualitative distinction between the GEM-4 and the PSM reflects the sensitive dependence of isostructural transitions on the details and convexity of the interaction potential~\cite{hemmer:2001}. The relatively long tail of the GEM-4  interaction indeed results in an additional energetic contribution to the lattice constant, and overcomes the entropy of mixing at low enough $T$. For GEM-$n$ with  $4<n<\infty$, one expects the threshold lattice constant difference between lattices with occupancy $n_{\mathrm{c}}$ and $n_{\mathrm{c}}+1$ to also depend on $T$ via the entropy of mixing. Qualitatively, we expect the isostructural critical point to  steadily decrease with the increase of $n$ and $n_\mathrm{c}$. A better knowledge of the drive for binary hard sphere mixtures of different size ratios to demix would be needed to estimate the equivalent behavior in the GEM-$n$.



With the equilibrium occupancy at hand, we study how the number of particles on a given site fluctuates in various $T$ regimes. This quantity is particularly important in determining which of the basic theoretical approximation is most analytically reasonable. Overall the relative fluctuation decreases with increasing temperature and density, but this progression is not monotonic (Fig.~\ref{fig:nc} Right). At low temperatures ($T<0.1$), a system with an equilibrium occupancy intermediate between two integers $\lfloor n_{\mathrm{c}}^{\mathrm{eq}}\rfloor <n_{\mathrm{c}}^{\mathrm{eq}}<\lceil n_\mathrm{c}^{\mathrm{eq}}\rceil$ can be seen as a random mixture of sites with occupancy $\lfloor n_{\mathrm{c}}^{\mathrm{eq}}\rfloor$ and occupancy $\lceil n_\mathrm{c}^{\mathrm{eq}}\rceil$. Excitations creating $\lceil n_\mathrm{c}^{\mathrm{eq}}\rceil+1$ or $\lfloor n_{\mathrm{c}}^{\mathrm{eq}}\rfloor -1$ defects are strongly suppressed. In that limit, the problem can be solved analytically. The probability of a site having $\lfloor n_{\mathrm{c}}^{\mathrm{eq}}\rfloor $ ($\lceil n_\mathrm{c}^{\mathrm{eq}}\rceil$) particles is  $n_{\mathrm{c}}^{\mathrm{eq}}-\lfloor n_{\mathrm{c}}^{\mathrm{eq}}\rfloor $ ($\lceil n_\mathrm{c}^{\mathrm{eq}}\rceil-n_{\mathrm{c}}^{\mathrm{eq}}$), and the resulting variance is $\sigma^2_{n_{\mathrm{c}}}=(2\lfloor n_{\mathrm{c}}^{\mathrm{eq}}\rfloor+1)n_{\mathrm{c}}^{\mathrm{eq}}-\lfloor n_{\mathrm{c}}^{\mathrm{eq}}\rfloor\lceil n_\mathrm{c}^{\mathrm{eq}}\rceil-(n_{\mathrm{c}}^{\mathrm{eq}})^2$. The $T=0.05$ results agree very well with this approximation (Fig.~\ref{fig:nc}(c)). This behavior explains why the mean-field cell model works increasingly well with decreasing $T$ and nearly quantitatively for $T\lesssim0.03$. As temperature increases, fluctuations become more pronounced, and vary continuously with occupancy. In this limit the crystal behavior is better approximated by a DFT-type treatment in which the free energy is optimized with respect to occupancy and fluctuations, assuming a Gaussian distribution~\cite{falkinger:2007}.

\begin{figure}
\includegraphics[width=3.5in]{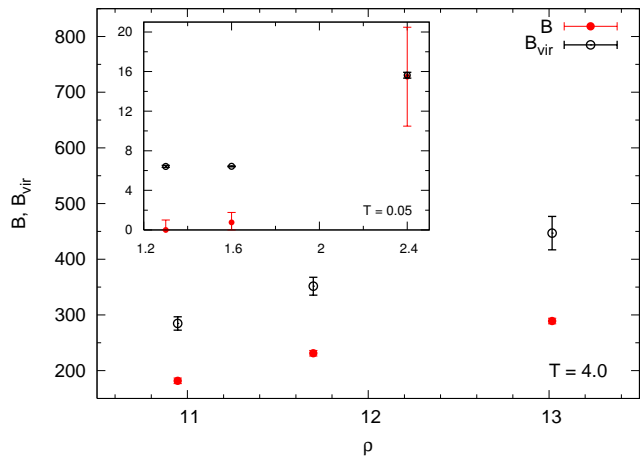}
\caption{The bulk modulus $B$ and its virial contribution $B_{\mathrm{vir}}$ (Eq.~\ref{eq:kappa}) at $T=4$ and $T=0.05$ (inset). The virial part $B_{\mathrm{vir}}$ is significantly softened by the cluster contribution, except in the low $T$ plateau regime ($\rho=2.4$).} \label{fig:bulkmodulus}
\end{figure}

\begin{figure*}
\includegraphics[width=3.2in]{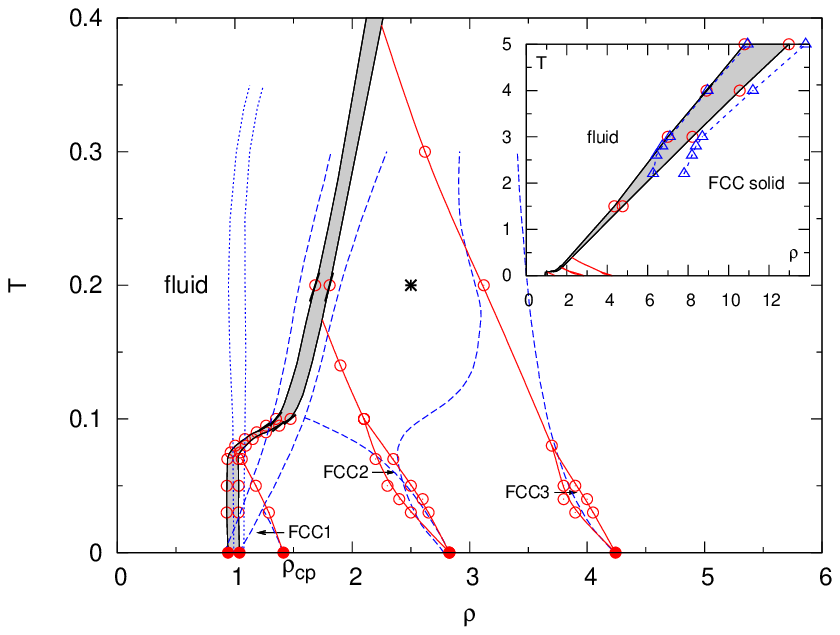} 
\includegraphics[width=3.2in]{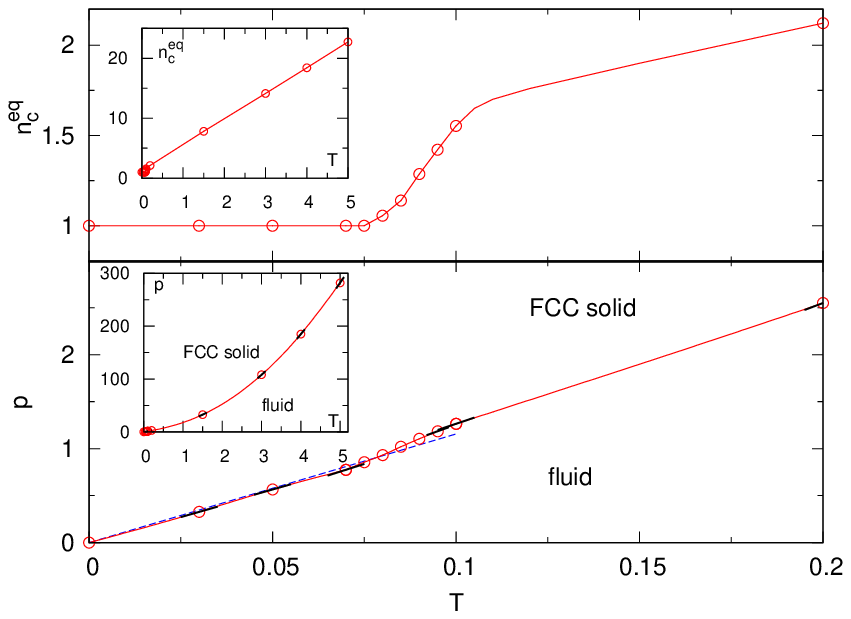}
\caption{Left: Simulation $T$-$\rho$ phase diagram obtained from the coexistence points  ($\odot$) and slopes (short black solid line). The other solid lines are guides for the eye, and the fluid-solid coexistence is shaded in grey. The errors are smaller than the symbol size. The low $T$ scenarios offered by the cell model~\cite{likos:1998} (dashed line) and  DFT~\cite{schmidt:1999} (dotted line) are shown for comparison. The boundaries of integer occupancy crystal phases (red solid line) terminate at $T=0$ results ($\bullet$) and compare nicely with the cell model for $T\leq0.03$, but not above. The $*$ region is discussed in the text. The agreement of the high $T$ results with DFT~\cite{falkinger:2007} ($\triangle$) break down for $T<3.0$ (inset). Top Right:  $n_{\mathrm{c}}^{\mathrm{eq}}$ at melting shows a rapid crossover between $0.07\lesssim T \lesssim0.1$ that connects the single-occupancy regime at low $T$ and the linear growth regime at high $T$ (inset).  Bottom Right: The corresponding crossover in the $p$-$T$ coexistence line separates the high temperature regime (inset) and the hard sphere-like regime. In this last region the coexistence pressure closely follows the hard sphere value $p_{\mathrm{coex}}=11.576 T$~\cite{zykova-timan:2010} (dashed line).} \label{fig:phasediagram}
\end{figure*}

As determined in earlier studies~\cite{mladek:2007,zhang:2010b}, the bulk modulus of cluster crystals, once decomposed into its two constituents (Sect.~\ref{subsect:response}), shows a strong softening correction due to the creation/annihilation process of lattice sites at high temperatures (Fig.~\ref{fig:bulkmodulus}). But in the integer occupancy regime, such as at $T=0.05$ for $\rho=2.4$, the virial part is the full resistance to external compression~\cite{zhang:2010b}. The two contributions to the coefficient of thermal expansion can be similarly calculated. At $T=4$ and $p=206$, for instance,
 calculations of the different components of Eq.~\ref{eq:alpha} agree with each other and with the decomposition (Table~\ref{table:alpha}).

\begin{table}
\caption{Calculation of $\alpha$ at $T = 4.0$ and $p=206$ with $n_{\mathrm{c}}^{\mathrm{eq}}=19.77$ using $(a1)$ a fit of the $T=3.0$-5.0 results (Fig.~\ref{fig:nc}), $(a2)$ multiple $NpT$ simulations with $n_{\mathrm{c}}=19.77$, $(a3)$ multiple $NpT$ simulations with $n_{\mathrm{c}}^{\mathrm{eq}}$ determined in $(a1)$, $(b1)$ histogram reweighting (Fig.~\ref{fig:reweight}), $(b2)$ histogram reweighting (Eq.~\ref{eq:mcvolumeT}),  $(c)$ a direct $[N]pT$ simulation (Eq.~\ref{eq:mcvolumep0}), and $(d)$ the fluctuation identity (Eq.~\ref{eq:alphafluc}).}
\begin{tabular}{c c c c}
\hline
$\alpha$ & $\alpha_{\mathrm{vir}}$ & $\left(\frac{\partial n_{\mathrm{c}}^{\mathrm{eq}}}{\partial T}\right)$ & $\frac{1}{v}\left(\frac{\partial v}{\partial n_{\mathrm{c}}}\right)$ \\
\hline
$0.195(10)^{a3}$ & $0.169(10)^{a2}$ & $-0.97(10)^{a1}$ & \\
$0.202(10)^{b1+b2+c}$ & $0.170(10)^{b2}$ & $-1.22(10)^{b1}$ & $-0.0265(3)^{c}$\\
 & $0.168(10)^{d}$ &  &\\
\hline
\end{tabular}
\label{table:alpha}
\end{table}

\subsection{Phase Diagram}


The fluid-crystal transition, which is the only  phase transition in the PSM, is extracted from the equilibrium free energies of both phases. The high and low $T$ regimes are treated separately, because the underlying physics is markedly different above and below the onset of clustering at coexistence around $T\sim 0.1$.

At high $T$, the coexistence regime on the $T$-$\rho$ phase diagram and the coexistence pressure on the $p$-$T$ phase diagram decreases smoothly towards $T\sim 0.1$. 
At melting $n_{\mathrm{c}}^{\mathrm{eq}}(\rho_{\mathrm{coex}})$ increases roughly linearly with $T$, because the coexistence density itself increases roughly linearly with $T$, as discussed above (Fig.~\ref{fig:phasediagram}). 
DFT results capture the transition relatively well for $T\geq3$, but strongly deviate below that point~\cite{falkinger:2007}. Yet attributing the deviation to a stronger hard-sphere character below that $T$~\cite{falkinger:2007}, is not supported by the numerical results, so another physical assumption of the theory may then be breaking down.

For $T\lesssim0.1$ at coexistence, two conflicting descriptions have previously been suggested (Fig.~\ref{fig:phasediagram}). First, a mixed DFT (fluid) and cell theory (crystal) study predicted that the fluid-solid transition
continuously approaches the HS limit upon lowering $T$~\cite{likos:1998}. Although the solid free energy was found to be indistinguishable from the HS results at finite
temperatures, changes to the fluid phase resulted in a steady drift away from the HS behavior for $T>0$. Second, a full DFT treatment predicted a crossover temperature
from a HS-like coexistence to a different regime around $T=0.35$~\cite{schmidt:1999}. We find the second scenario to be qualitatively correct, although its
predicted crossover temperature is nearly three times larger than the numerically determined one. For $T\sim 0.1$, the coexistence crystal density gets smaller than the HS
close packing density $\rho_{\mathrm{cp}}=\sqrt{2}$. Clustering then gets suppressed and the fluid-solid coexistence curve is inflected. Below $T<0.07$, the
fluid coexists with an essentially singly-occupied FCC phase. The transition is hard sphere-like with the corresponding coexistence pressure and densities (Fig.~\ref{fig:phasediagram}).  

That there should be a crossover can physically be understood from the fact that at $T\ll 1$ particle overlaps are rare and uncorrelated. They therefore marginally
stabilize both the fluid and crystal phases in a similar way. Once the concentration of overlaps becomes sufficiently high, however, the difference in structure between the ordered and disordered
phases matters. This distinction between the high and low $T$ regime also has a dynamical signature~\cite{santos:2005,suh:2010}. Collisions between particles can be divided
in two types: soft refractive collisions, in which a particle goes through another, and hard reflective collisions, in which particles elastically bounce back from
each other. At temperatures $T\lesssim0.3$ from molecular dynamics simulations~\cite{suh:2010} or $T\lesssim0.25-0.5$ from an Enskog-type theoretical analysis~\cite{santos:2005},
the first collision type is highly suppressed because the particle momenta are low, and the collision frequency of the second type is as high as in hard spheres. 


Unlike for the GEM-4 model, where first-order phase transitions between integer-occupancy lattices are observed at low $T$, crystal lattice occupancy in the PSM changes continuously. The stability regime of integer-occupancy phases, such as FCC2 and FCC3, only occupies a narrow slice of the phase diagram at low temperatures. Above $T\sim 0.1$, these phases join the continuum of non-integer-occupancy crystals whose $T$-$\rho$ stability regime is but a line that extends smoothly up to the liquid-solid coexistence boundary. The cell model fails to capture this regime's behavior (Fig.~\ref{fig:phasediagram}). It notably erroneously predict the existence of an intermediate single-double-triple occupancy regime ($*$ in Fig.~\ref{fig:phasediagram}). The limited extent of occupancy fluctuations at this low $T$ (Fig.~\ref{fig:nc}(c)) shows that such high-energy excitations are unphysical. Simulations indeed clearly delineate the single-double from the double-triple occupancy regimes.

\section{Conclusion}
In this paper, we have reformulated the extended thermodynamics of cluster crystals such that it lends itself to 
 $[N]pT$ ensemble simulations~\cite{orkoulas:2009}. Using this approach and thermodynamic integration, we efficiently determined the phase diagram of the canonical penetrable sphere model, which forms cluster crystals at high densities.  The $[N]pT$ ensemble approach naturally allows for histogram reweighting, which was here used to calculate the response functions, and could be particularly useful for studying the critical properties of cluster crystal formers, such as the GEM-4. 

The resulting formalism and method are analogous to a constant pressure version of the approach used for determining the binary hard sphere mixture phase diagram~\cite{kranendonk:1991a,kranendonk:1991b}, which it should be able to calculate with some adjustments. The approach also allows for the study of models that form crystals in which the vacancy concentration is relatively large, such as hard cubes~\cite{smallenburg:2011,frenkel:2011}. The key limitation for these implementations is the need for an efficient particle insertion algorithm and a good initial guess of the conjugate field.


\begin{acknowledgments}
We gratefully thank N. Wilding for numerous discussions and particularly for the key suggestion that the $[N]pT$ approach may be useful to the study of cluster crystals. We also thank D. Frenkel, R. Jack, and B. Mladek for discussions at various stages of this project. We acknowledge National Science Foundation Grant No.~NSF DMR-1055586.
\end{acknowledgments}

\appendix


\begin{thebibliography}{58}
\bibitem[{foo({\natexlab{b}})}]{footnote:2}
\bibinfo{note}{The differential can be obtained for any $n_{\mathrm{c}}$, but
  only has a clear-cut thermodynamic interpretation when its value is
  calculated for the equilibrium cluster occupancy.}
\end{thebibliography}

\end{document}